\documentstyle[12pt,here]{article}
\def\beq{\begin{equation}}
\def\eeq{\end{equation}}
\def\bea{\begin{eqnarray}}
\def\eea{\end{eqnarray}}
\def\bq{\begin{quote}}
\def\eq{\end{quote}}

\newcommand{\hl}{\hline}

\newcommand{\bo}{$B^0$}
\newcommand{\bb}{$\bar{B}^0$}
\newcommand{\ko}{$K^0$}
\newcommand{\kb}{$\bar{K}^0$}
\newcommand{\ks}{$K_S$}
\newcommand{\kl}{$K_L$}
\newcommand{\dg}{$\Delta \Gamma$}
\newcommand{\ol}{${\cal O}(\lambda^3)$}
\newcommand{\ie}{${\rm Im}(\varepsilon)$}
\newcommand{\rd}{${\rm Re}(\delta)$}
\newcommand{\jk}{$J/\Psi \  K_S$}
\newcommand{\jkl}{$J/\Psi \  K_L$}

\def\cp1sub{\setlength{\unitlength}{8pt}\begin{picture}(2,1)
\mbox{\scriptsize CP} \end{picture}}

\def\gappeq{\mathrel{\rlap {\raise.5ex\hbox{$>$}}
{\lower.5ex\hbox{$\sim$}}}}

\def\lappeq{\mathrel{\rlap{\raise.5ex\hbox{$<$}}
{\lower.5ex\hbox{$\sim$}}}}

\begin{document}
\pagestyle{empty}
\begin{flushright}
CERN-TH/99-244
\end{flushright}
\vspace*{5mm}
\begin{center}
{\bf CP, T AND CPT VERSUS TEMPORAL ASYMMETRIES}\\
{\bf FOR ENTANGLED STATES OF THE $B_d$-SYSTEM}
\\
\vspace*{1cm}
{\bf  M.C. Ba\~ nuls} \\
\vspace*{0.3cm}
IFIC, Centro Mixto Univ. Valencia - CSIC, E-46100 Burjassot (Valencia), Spain\\
{\tt banuls@goya.ific.uv.es}\\
\vspace*{0.3cm}
and \\
\vspace*{0.3cm}
{\bf J. Bernab\'eu}\footnote{On leave from
\vspace*{0.3cm}
Departamento de F\'{\i}sica Te\'orica, Univ. Valencia, E-46100
Burjassot (Valencia), Spain.} \\
\vspace{0.3cm}
Theoretical Physics Division, CERN \\
CH - 1211 Geneva 23 \\
{\tt bernabeu@mail.cern.ch}\\
\vspace*{2cm}
{\bf ABSTRACT} \\ \end{center}
\vspace*{5mm}

The observables used in the $K$-system to characterize T and CPT violation
are no longer useful for the $B_d$-system, since the
 width difference
\dg~between the physical states is vanishingly small.  We show that only
\ie~and \rd~can survive
if $\Delta \Gamma=0$,
and build alternative CP-odd, CPT-odd, T-odd and temporal asymmetries for
the \mbox{$B_{\pm}
\rightarrow B^0, \, \bar{B}^0$} transitions. These quantities enable us
to test T and CPT
invariances of the effective Hamiltonian for the $B$-system.
The method needs the CP eigenstates
$B_{\pm}$, which can be tagged unambiguously to \ol~from the
entangled states of a $B$-factory.

\vspace*{2cm}
\begin{flushleft}
CERN-TH/99-244\\
August 1999
\end{flushleft}

\vfill\eject

\setcounter{page}{1}
\pagestyle{plain}
The time evolution of a neutral-meson system is governed by
an effective Hamiltonian.
The study of its symmetries corresponds to the problem of mixing,
excluding effects from direct decays.
The non-invariance of this Hamiltonian is termed as indirect violation.
In the case of the neutral $K$-system, such a study has been carried out
by the CP-LEAR experiment~\cite{cplear}
for CP-, T-violating and CP-, CPT-violating observables
constructed from the preparation of definite flavour states \ko-\kb.
After time evolution of these tagged mesons, the semileptonic
decay projects again on a definite flavour state.
Although some doubts~\cite{ag98} remain associated with possible
CPT violation in the semileptonic decays,
the experimental results are interpreted in terms of non-invariance
of the mixing Hamiltonian.
However, the observables for both T-odd and CPT-odd quantities do  not need
 only a fundamental violation of these symmetries but also
non-vanishing absorptive components in the effective Hamiltonian.
In the case of the neutral $K$-system, such an ingredient is provided by
the different lifetimes of the physical states with definite mass,
\ks~and \kl.
This study when applied to entangled states of \ko-\kb, as
produced by $\Phi$ decay at DAPHNE, allows~\cite{bu92} additional
separation of the different T-odd and CPT-odd asymmetries
in mixing and decay.

None of the T-odd and CPT-odd observables based on flavour tag and
discussed for
the $K$-system is useful for the $B_d$ case.
In this system, one expects~\cite{kh87} a negligible value of the
width difference \dg~between the physical states.
As a consequence, the observables based on flavour tags vanish
even if there is a fundamental violation of the symmetries.
But the $B_d$ entangled states obtained in the $B$ factories
from the $\Upsilon(4 S)$ decay can be used for appropriate alternative tags.
In this letter we propose the construction of alternative asymmetries
in order to test CP, T and CPT invariances.
To this end we study the possibilities offered by CP eigenstates
of the $B_d$-system.
These states can be identified~\cite{bb99} unambiguously to \ol,
which is sufficient to discuss both CP-conserving and CP-violating
amplitudes in the effective Hamiltonian for $B_d$ mesons.
Here $\lambda$ is the flavour-mixing parameter of the CKM matrix~\cite{ko73}.
We show that the time-dependent CP-, T-, and CPT-odd asymmetries for
the semileptonic decay of the CP-tagged state contain fundamental
information on the CP-, T-violating parameter $\varepsilon$ and the
CP-, CPT-violating parameter $\delta$ in the effective Hamiltonian.
More precisely, when $\Delta \Gamma =0$, the parameters signalling
the violation of these symmetries are \ie~and \rd.
We also consider the temporal asymmetry, defined from the inversion
in the order of appearance of the final decay products from the
entangled state.
The main goal of this work is thus the connection of the four
CP-odd, CPT-odd, T-odd and temporal asymmetries for the transitions
\mbox{$B_{\pm} \rightarrow B^0, \, \bar{B}^0$} with the parameters
\rd~and \ie, which quantify CP-, CPT-violation and CP-, T-violation,
respectively, in the effective Hamiltonian.

The flavour states are connected by \mbox{${\rm CP} |B^0 \rangle = {\rm
CP}_{12}^*
|\bar{B}^0
\rangle$}.  Contrary to the corresponding phase of CPT operation, which is
rephasing-invariant, the phase $C\!P_{12}$ is parametrization-variant.
We keep track of its presence.
The corresponding CP eigenstates are thus
\beq
|B_{\pm} \rangle = \frac{1}{\sqrt{2}} (I \pm {\rm CP} ) |B^0 \rangle=
\frac{1}{\sqrt{2}} ( |B^0 \rangle \pm {\rm CP}_{12}^* |\bar{B}^0 \rangle)~.
\eeq

These states are well defined iff the CP operator is unique.
The identification of the CP operator needs the explicit separation
of the Hamiltonian into CP-conserving and CP-violating components.
Its determination is based~\cite{bb99} on the requirement of CP
conservation, to \ol, in the $(sd)$ and $(bs)$ sectors.
This fixes, up to a common rephasing of all the quark fields,
the CP phases $(\theta_s-\theta_d)$ and $(\theta_b-\theta_s)$,
respectively.
Thanks to the cyclic relation between CP phases,
\beq
e^{i(\theta_b-\theta_d)}=e^{i(\theta_b-\theta_s)}e^{i(\theta_s-\theta_d)}
\eeq
is also fixed.
Since in the $(bd)$ sector both CP-conserving and CP-violating amplitudes
are present at \ol, this fixing allows us to classify CP violation in
$B_d$ decays by referring it to this well defined CP-conserving direction.
In this sense, a $B_d$ decay that is governed by the couplings of the
$(sd)$ or $(bs)$ unitarity triangles, or by the $V_{cd} V_{cb}^*$ side
of the $(bd)$ triangle, will not show any CP violation to \ol.
We may say that such a channel is free from direct CP violation.

The physical states $B_{1,2}$ of definite mass and lifetime  are written
in terms of the CP eigenstates $B_{\pm}$ as
\bea
|B_1 \rangle &=& \frac{1}{\sqrt{2 (1+|\varepsilon_1|^2)}} \left [|B_+
\rangle +
\varepsilon_1 |B_- \rangle \right ]
\nonumber \\
|B_2 \rangle &=& \frac{1}{\sqrt{2 (1+|\varepsilon_2|^2)}} \left [|B_-
\rangle +
\varepsilon_2 |B_+ \rangle \right ] ~.
\label{fis}
\eea
The complex parameters $\varepsilon_{1,2}$ describe the CP mixing
in the corresponding physical state.
Contrary to other definitions of these parameters found in the literature,
ours is rephasing-invariant~\cite{bb98} without invoking the introduction
of a particular decay channel.
Being independent of the parametrization, $\varepsilon_1$ and
$\varepsilon_2$ are physical quantities iff the CP operator is
well defined as discussed above.

If CPT is a good symmetry, $\varepsilon_2=\varepsilon_1$.
We thus  may alternatively use the parameters $\varepsilon$ and $\delta$,
which admit a simpler interpretation in terms of symmetries
\beq
\varepsilon \equiv \frac{\varepsilon_1 + \varepsilon_2}{2}, \hspace{2cm}
 \delta \equiv \varepsilon_1 - \varepsilon_2~.
\eeq

The states (\ref{fis}) or $(\varepsilon, \, \delta)$ are obtained
from the diagonalization of the non-Hermitian effective
Hamiltonian $H=M-\frac{i}{2}\Gamma$, with $M^+=M$, $\Gamma^+=\Gamma$.
In the limit $\Delta \Gamma=0$~\cite{kh87}, the anti-Hermitian part of $H$
is proportional to the identity matrix, and so $H$ can be diagonalized
by a unitary transformation, and its physical states will be orthogonal.
Explicitly, assuming that CPT violation is small and we may neglect
terms that are quadratic in $\Delta \equiv H_{22}-H_{11}$, we obtain
(for $\Delta \Gamma =0$) the parameters
\bea
& &{\rm Re}(\varepsilon)=0, \hspace{2cm} \frac{{\rm Im}(\varepsilon)}
{1+|\varepsilon|^2}=\frac{{\rm Im}(M_{12} {\rm CP}_{12}^*)}{\Delta m}
\nonumber \\
& &\frac{{\rm Re}(\delta)}{1+|\varepsilon|^2}=\frac{\Delta}{\Delta m},
\hspace{2cm} {\rm Im}(\delta)=0
\label{parab}
\eea

As can be seen, $\varepsilon$ becomes purely imaginary and $\delta$ real
in the limit of
negligible $\Gamma_{12}$, in which \mbox{$\Delta \Gamma \ll \Delta m$}. We then
find $\varepsilon_2=-\varepsilon_1^*$ and the orthogonality of the
states~(\ref{fis})
is apparent.

The restrictions imposed by each symmetry are the following:
\begin{itemize}
\item{CP conservation imposes ${\rm Im}(M_{12} {\rm CP}_{12}^*)=0$ and
$H_{11}=H_{22}$;
\item{CPT invariance requires $H_{11}=H_{22}$}};
\item{T invariance imposes ${\rm Im}(M_{12} {\rm CP}_{12}^*)=0$}.
\end{itemize}
As a consequence, CPT invariance leads to $\Delta=0$ and thus $\delta=0$,
irrespective
of the value of $\varepsilon$. Similarly, T invariance leads to
$\varepsilon=0$,
independently of the value of $\delta$. CP conservation requires both
$\varepsilon=\delta=0$. Therefore,
\begin{itemize}
\item{${\rm Im}(\varepsilon) \neq 0$ indicates the presence of both CP and
T violation;}
\item{${\rm Re}(\delta) \neq 0$ means that CP and CPT violation exist}.
\end{itemize}

In a $B$ factory, charge conjugation and Bose statistics require
that the \bo-\bb~state produced in the $\Upsilon(4 S)$ decay be given by
\beq
|\Psi_{in}\rangle = \frac{1}{\sqrt{2}} \left ( | B^0 (\vec{k}),\bar{B}^0
(-\vec{k}) \rangle - | \bar{B}^0 ( \vec{k}), B^0 (-\vec{k}) \rangle \right )~.
\label{initial}
\eeq

The correlation implied by Eq. (\ref{initial}) remains at any moment
in the evolution of the system and therefore, the decay product observed
first in one side of the detector allows the tagging of the meson in the
other side
 at the moment the decay took place.
This correlation holds not only for the flavour basis
$(B^0,\, \bar{B}^0)$, but also for CP eigenstates $(B_+,\, B_-)$
and for physical states $(B_1,\, B_2)$, independently
of the values of $\varepsilon$ and $\delta$.

We will use the notation $(X, \, Y)$ for the final state.
Here $X$ is the decay product observed
with momentum $\vec{k}$ at a time $t_0$,
whereas $Y$ is detected with momentum $-\vec{k}$ at a later time $t$.
We will describe the process in terms of
the time variables $\Delta t=t-t_0$ and $t'=t_0+t$.
The probability of
finding the arbitrary final state $(X,Y)$
from the initial state (\ref{initial})
may be written as:
\bea
|( X,Y)|^2 &=& \frac{1}{16}
\left | \frac{1+ \varepsilon_2}{1+ \varepsilon_1} \right |^2
|\langle X | B_1\rangle |^2 |\langle Y | B_1\rangle |^2 e^{-{\Gamma} t'} 
\times \nonumber \\
& & \phantom {\times []} \times
\left \{
(\eta_+ + \eta_-) - (\eta_+ - \eta_-) \cos (\Delta m \Delta t)
-2 \eta_{\rm im} \sin (\Delta m \Delta t)
\right \}~,
\label{prob}
\eea
 $\Gamma$ being the common width of $B_{1,2}$.
Since \dg $=0$, in this and the following expressions one should take
$\varepsilon_2=-\varepsilon_1^*$.
The $\eta$ coefficients are defined according to
\beq
\eta_+=|\eta_X + \eta_Y|^2, \hspace{1cm}
\eta_- = |\eta_X - \eta_Y|^2 , \hspace{1cm}
\eta_{\rm im} = {\rm Im} [(\eta_X +\eta_Y)(\eta_X^* - \eta_Y^*)]~,
\label{coef}
\eeq
with
\beq
\eta_X \equiv \frac{ \langle X |B^0 \rangle -
\frac{1-\varepsilon_2}{1+\varepsilon_2} {\rm CP}_{12}^*\langle X |\bar{B}^0
\rangle}
{\langle X |B^0 \rangle + \frac{1-\varepsilon_1}{1+\varepsilon_1} {\rm
CP}_{12}^*\langle
X |\bar{B}^0 \rangle}=
\frac{1+\varepsilon_1}{1+\varepsilon_2}\cdot
\frac{\varepsilon_2  \langle X |B_+\rangle+\langle X |B_-\rangle}
{\langle X |B_+\rangle+\varepsilon_1 \langle X |B_-\rangle}~,
\label{eta}
\eeq
and an analogous expression for $\eta_Y$.
We have written Eq. (\ref{eta}) in terms of both flavour and
CP eigenstates, in view of the tags we propose below.
One can check that, for $Y=X$, only $\eta_+$ remains and the
probability $|(X,X)|^2$ vanishes for $\Delta t=0$.
This is a kind of EPR correlation, imposed here by Bose statistics~\cite{li68}.

The integration of the probability (\ref{prob}) over $t'$ between $\Delta t$
(always positive with our definition) and $\infty$ gives the intensity for
the chosen final state, depending only on $\Delta t$
\beq
I(X, \, Y; \, \Delta t)= \int_{\Delta t}^{\infty} d t' |(X, \, Y)|^2~.
\eeq

To perform a CP tag, let $X$ be a CP eigenstate produced along
the CP-conserving direction.
Since, as discussed above, such a decay is free from direct CP violation,
this assures that at $t=t_0$ the living meson of the other side
had the opposite CP eigenvalue.
Among others, one example of a decay channel with the properties
we are looking for is given by $X \equiv$ \jk, with CP$=-$, or by
$X \equiv J/\Psi \ K_L$, with CP$=+$, governed by the ``CP-allowed''
side $V_{cb}V_{cs}^*$ of the $(bs)$ triangle.
Then, the detection of such a final state leads to the preparation of the
remaining $B_d$ meson in the complementary CP eigenstate to \ol.

With these considerations, let us assume that, at $t=t_0$, the $B_d$ meson
is prepared as a $B_+$.
After this CP tag, the time evolution acts during $\Delta t$ and we ask
for the probability of its transition to \bo.
In our result (\ref{prob}) for the probability from the entangled state,
this $B_+ \stackrel{\Delta t}{\rightarrow} B^0$ transition corresponds to
a final state $X=$ \jk, $Y=\ell^+$.
In Table~\ref{tab:1} we present the transitions of $B_d$ states,
connected to the original one by CP-, CPT- and T- symmetry transformations.
The last column, showing $\Delta t$, is obtained by inverting
 the order of appearance of the decay products.
The second row in the table shows the $(X, \, Y)$ characterization
for each one of these processes, i.e., the configuration of the final
decay products, which will signal such a transition at the meson level.

\begin{table}[h]
\begin{center}
\begin{tabular}{|c|c|c|c|c|} \hl
\begin{minipage}{3cm}
\begin{center}
$B_+ \stackrel{\phantom{-}}{\rightarrow} B^0$
\hspace*{-.2cm} \rule[10pt]{3cm}{.1pt}
\\
\vspace*{-14pt}
$\phantom{\stackrel{-}{-}}
(J/\Psi \ K_S, \, \ell^+)
\phantom{\stackrel{-}{-}}$
\end{center}
\end{minipage}
& CP & CPT & T & $\Delta t$ \\ \hl
Transition & $B_+ \stackrel{\phantom{-}}{\rightarrow} \bar{B}^0$ &
$\bar{B}^0 \stackrel{\phantom{-}}{\rightarrow} B_+$ &
$B^0 \stackrel{\phantom{-}}{\rightarrow} B_+$ & $\bar{B}^0
\stackrel{\phantom{-}}{\rightarrow} B_-$ \\ \hl
$\phantom{\stackrel{-}{-}}
(X,\, Y)
\phantom{\stackrel{-}{-}}$ & $($\jk, $\ell^-)$ & $(\ell^+$, \jkl $)$ &
$(\ell^-$, \jkl $)$ & $(\ell^+$, \jk $)$ \\
\hl
\end{tabular}
\caption{Transitions and $(X,\, Y)$ final decay products connected to $B_+
 \rightarrow B^0$ and $( J/\Psi \ K_S, \, \ell^+)$ by the four
transformations.}
\label{tab:1}
\end{center}
\end{table}

From the five different processes indicated in Table~\ref{tab:1}, we
construct four
intensity asymmetries
\beq
A(X,\, Y)=\frac{I(X,\,Y)-I(J/\Psi \ K_S,\, \ell^+)}{I(X,\,Y)+I(J/\Psi \
K_S,\, \ell^+)}
\eeq
as a function of $\Delta t$.
They are odd under the four transformations considered.

To first order in the CPT parameter $\delta$, and in the limit \dg $=0$,
we obtain the following results:
\begin{itemize}
\item
The CP-odd asymmetry gives
\beq
A_{{\rm CP}}\equiv A(J/\Psi \ K_S, \, \ell^-)=
-2 \frac{{\rm Im} (\varepsilon)}{1+|\varepsilon|^2} \sin (\Delta m \Delta t)
+\frac{1-|\varepsilon|^2}{1+|\varepsilon|^2} \frac{2{\rm
Re}(\delta)}{1+|\varepsilon|^2} \sin^2 \left (\frac{\Delta m \Delta t}{2}
\right )
\label{acp}
\eeq
with contributions from both T-violating ($\Delta t$ odd) and CPT-violating
($\Delta t$ even) terms.
\item
The CPT-odd asymmetry yields
\beq
A_{{\rm CPT}} \equiv A(\ell^+, \, J/\Psi \ K_L)  =
\frac{1-|\varepsilon|^2}{1+|\varepsilon|^2} \frac{2 {\rm
Re}(\delta)}{1+|\varepsilon|^2}
\frac{1}{1-2 \frac{{\rm Im} (\varepsilon)}{1+|\varepsilon|^2} \sin (\Delta
m \Delta t)}
\sin^2 \left (\frac{\Delta m \Delta t}{2}\right )
\label{acpt}
\eeq
with a needed contribution from $\delta \neq 0$, including T-even and T-odd
terms.
\item
The T-odd asymmetry gives
\beq
A_{\rm T} \equiv A(\ell^-, \, J/\Psi \ K_L) =
-2 \frac{{\rm Im} (\varepsilon)}{1+|\varepsilon|^2} \sin (\Delta m \Delta t)
\left[ 1 - \frac{1-|\varepsilon|^2}{1+|\varepsilon|^2}~~ \frac{2 {\rm
Re}(\delta)}{1+|\varepsilon|^2} ~~\sin^2 \left (\frac{\Delta m \Delta t}{2}
\right)\right]
\label{aT}
\eeq
which needs $\varepsilon \neq 0$, including CPT-even and CPT-odd terms.
\item
Lastly, the temporal asymmetry satisfies the equality
\beq
A_{\Delta t} \equiv A(\ell^+, \, J/\Psi \ K_S)=
  A(\ell^-, \, J/\Psi \ K_L)\equiv A_{\rm T}~.
\label{adt}
\eeq
\end{itemize}

It is worth noting that, contrary to the other asymmetries, $A_{\Delta t}$
is not associated to the transformation of the original transition
under a fundamental symmetry.
Nevertheless, it may provide information on the symmetry properties of the
system.
The experimental $\Delta t$ asymmetry is different from the T-odd asymmetry.
In our case, the equality (\ref{adt}) is a consequence of \dg $=0$
and it can be used as a consistency test of this assumption.
In general, the equality of the probabilities for the T-inverted and
$\Delta t$-inverted processes is only valid for Hermitian Hamiltonians,
up to a global (proportional to unity) absorptive part.
It is not satisfied, for instance, for the $K$-system.
On the contrary, for $B_d$-mesons the T-odd asymmetry (\ref{aT})
becomes an odd function of time, once \dg $=0$ is assumed.

The results (\ref{acp})-(\ref{adt}) can be used to test consistencies
and extract the parameters
\beq
\frac{2 {\rm Im} (\varepsilon)}{1+|\varepsilon|^2}
, \hspace{2cm}
\frac{1-|\varepsilon|^2}{1+|\varepsilon|^2} \frac{2 {\rm Re}(\delta)}
{1+|\varepsilon|^2},
\eeq
fundamental for CP, T violation and CP, CPT violation, respectively.
Their connection with the matrix elements of the effective Hamiltonian
is given in Eq. (\ref{parab}).

In the Standard Model, $\delta=0$, and the four asymmetries collapse to
\bea
A_{\rm CP~}&=&A_{\rm T}=A_{\Delta t}=
-2 \frac{{\rm Im} (\varepsilon)}{1+|\varepsilon|^2} \sin (\Delta m \Delta t)
\nonumber \\
A_{\rm CPT}&=&0
\eea
In terms of the angles of the $(bd)$ unitarity triangle, the CP-mixing
parameter is then given by
\bea
&&\frac{2{\rm Im} (\varepsilon)}{1+|\varepsilon|^2}=\sin (2 \beta)=-\frac{2
\eta (1-\rho)}{(1-\rho)^2+\eta^2}
\nonumber \\
&&\nonumber \\
&&\frac{1-|\varepsilon|^2}{1+|\varepsilon|^2}=\cos (2 \beta) =
\frac{(1-\rho)^2-\eta^2}{(1-\rho)^2+\eta^2}
\eea
where $(\rho, \, \eta)$ are the couplings appearing in the
Wolfenstein parametrization~\cite{wo83}. $A_{\rm CP}$ has been measured
\cite{cdf99} by
the CDF Collaboration at Fermilab, using a flavour tag, and the result
interpreted in
the Standard Model with $\sin(2\beta) = 0.79^{+0.41}_{-0.44}$. However, the
other three
asymmetries do not need only a flavour tag but also a CP tag, as shown in
Table 1.

To summarize, we have been able to bypass the vanishing of T-odd
and CPT-odd asymmetries in the limit $\Delta\Gamma =0$.
The crucial point consists in going beyond the flavour tags
used for the $K$-system.
The production of entangled states of $B_d$ mesons in a $B$-factory allows
an unambiguous tag of CP eigenstates.
This CP tag makes use of decay channels along the CP-conserving direction,
such as that illustrated by \jk.
Starting with the $B_+ \stackrel{\Delta t}{\rightarrow} B^0$ transition,
we have considered four asymmetries, which are odd under the CP, CPT and
T symmetry transformations and under the temporal inversion of the
decay products.
We prove  that, in the limit $\Delta\Gamma =0$, the T-odd asymmetry equals the
temporal asymmetry.
CP-odd and CPT-odd asymmetries contain, on the contrary, both $\Delta t$-odd
and $\Delta t$-even terms.
Whereas the CPT asymmetry needs $\delta \neq 0$, in the presence or absence
of  $\varepsilon$, the CP asymmetry can be generated by  a non-vanishing
value of either $\varepsilon$ or  $\delta$, or of both.
All these four asymmetries, experimentally different, provide a
consistent set of observables able to disentangle the fundamental
parameters $\varepsilon$ and $\delta$ of T-violation and CPT-violation,
respectively.

\section*{Acknowledgements}

The authors would like to acknowledge S. Bilenky, F. Botella, E. Gabathuler,
 C. Jarlskog, N. Pavlopoulos, E. de Rafael and A. Santamar\'{\i}a for
discussions on
the topic of this paper.
\mbox{M. C. B.} is indebted to the Spanish Ministry of Education and
Culture for her
fellowship, and to SISSA for hospitality. This research was supported by
CICYT, Spain,
under Grant AEN-96/1718.

\end{document}